# Alzheimer's Disease Diagnosis by Deep Learning Using MRI-Based Approaches

Sarasadat Foroughipoor[1], Kimia Moradi[1], Hamidreza Bolhasani[1*]

***Abstract***

The most frequent kind of dementia of the nervous system, Alzheimer's disease, weakens several brain processes (such as memory) and eventually results in death. The clinical study uses magnetic resonance imaging to diagnose AD. Deep learning algorithms are capable of pattern recognition and feature extraction from the inputted raw data. As early diagnosis and stage detection are the most crucial elements in enhancing patient care and treatment outcomes, deep learning algorithms for MRI images have recently allowed for diagnosing a medical condition at the beginning stage and identifying particular symptoms of Alzheimer's disease. As a result, we aimed to analyze five specific studies focused on AD diagnosis using MRI-based deep learning algorithms between 2021 and 2023 in this study. To completely illustrate the differences between these techniques and comprehend how deep learning algorithms function, we attempted to explore selected approaches in depth.

## 1. Introduction

Alzheimer's disease (AD) is the most widespread form of dementia, affecting millions worldwide [2]. AD is a chronic neurological brain disorder that leads to cognitive decline, loss of mental abilities, and other impairments [3]. Sadly, it is also the most common form of dementia globally [4]. It is estimated that by 2050, the number of individuals with dementia will reach 152 million, up from 82 million in 2030 [5]. AD is a progressive condition that causes the deterioration of brain tissue and neuronal cells, gradually decreasing cognitive function and memory. It also impacts a patient's ability to perform daily tasks, such as writing, speaking, and reading, and can cause difficulties recognizing loved ones [6]. It is eventually fatal.

Many individuals experience memory impairment, broadly categorized as Mild Cognitive Impairment (MCI). However, some may also face challenges with cognitive function that lead to dementia issues. Those with mild dementia often struggle to perform everyday tasks due to symptoms such as loss of memory, anxiety, changes in personality, feelings of loss, and difficulty carrying out daily tasks. As dementia progresses to the moderate stage, patients require additional help and support. Symptoms worsen, including amnesia and confusion, difficulty recognizing family and friends, significant personality changes, insomnia, and changes in behavior such as paranoia and irritability. In severe dementia, symptoms can become even more debilitating, leaving patients incapable of communicating and needing constant care. Basic tasks like holding their head up or sitting in a chair can become difficult, and loss of bladder control is common. Unfortunately, severe dementia can ultimately result in death [7].

To date, there is no known cure for this illness. Detecting it early is crucial for prompt treatment and to delay its progression [8, 9]. Accurate predictions of likely progression from MCI to AD are also necessary [10, 11]. MCI is recognized as the stage between age-related cognitive impairment and AD. Identifying individuals with MCI at high risk of developing AD is important for effective treatment [12]. Patients with

[1] Department of Computer Engineering, Science and Research Branch, Islamic Azad University, Tehran, Iran.
* Corresponding Author: hamidreza.bolhasani@srbiau.ac.ir

MCI can be classified as either progressive MCI (pMCI) or stable MCI (sMCI), indicating that the patient has not necessarily transitioned to AD [13].

Numerous neuroimaging techniques have been developed for studying the brain in recent years. Magnetic Resonance Imaging (MRI) is a method that utilizes magnetic fields and radio waves to produce a 3-dimensional image of the brain. Positron emission tomography (PET) is another method that maps the brain's regions using radioactive tracers, which can help identify protein plaques associated with dementia [14]. The MRI technique boasts numerous advantages in the field of medical imaging, including high imaging flexibility, exceptional tissue contrast, and the ability to provide valuable structural information about the human brain, all without subjecting patients to ionizing radiation [15].

Neuroimaging employs Machine Learning (ML) to enhance the precision of dementia subtype classification. To enable an ML algorithm, specific pre-processing procedures are necessary. The machine learning-based classification process encompasses various stages, such as feature extraction, feature selection, dimensionality reduction, and classification algorithms [16]. Integrating Deep Learning (DL) with non-invasive image-based Computer-Aided Diagnostic (CAD) systems for the early identification of illnesses has yielded remarkable performance improvements and substantial medical benefits over the years. Recently, significant advances have been made in MRI-based CAD systems, displaying immense potential in accurately detecting AD patients [17].

Convolutional Neural Networks (CNN) application in DL has yielded promising results in utilizing digital brain scans of patients diagnosed with AD as a viable means to assist clinicians in making informed medical decisions [18]. DL approaches based on deep CNN enable data-driven feature extraction directly from image data [19]. CNN is the method that is most frequently applied and depicted in research studies on processing and analyzing images of the brain [6]. CNN directly takes 2D or 3D pictures and automatically learns relevant global and higher-level local features, eradicating the many measurement mistakes caused by the traditional hand-crafted feature [6]. These techniques have exhibited superior performance compared to traditional approaches reliant on predetermined characteristics in the majority of image processing and computer vision tasks [20].

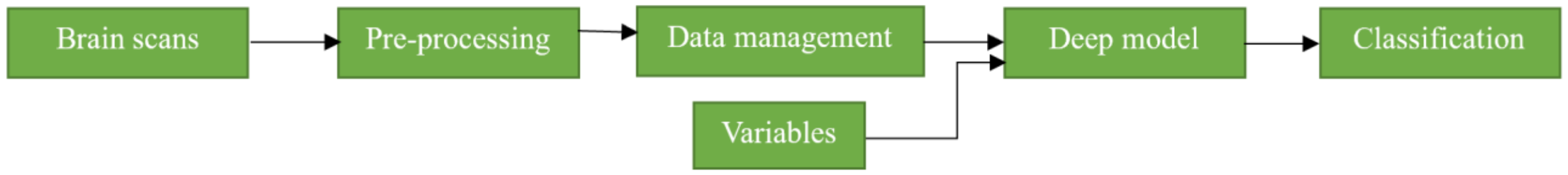

Figure. 1. An example of a computer-aided AD detection system's block diagram.

## 1-2. A brief look at the CNN and evaluation parameters

### 1-2-1. CNN

The following paragraphs will describe the CNN model approach for each group of layers.

The convolutional layer is a key component of a DL CNN. It serves as a foundational building block of the network, extracting features from input data. The output of this layer is referred to as feature maps, that are composed of sets of 2D matrices. By convolving the input image with a predetermined number of feature detectors, this layer extracts features that help the network recognize low-level properties of the images, such as edges, corners, and colors. During training, each filter learns to recognize these features through trial and error [35]. The resultant feature map, or response, is attained through the process of convolving the input image with weight filters and bias values prior to its transferal to successive layers [27].

The pooling layer is commonly employed after convolutional layers to sub-sample feature maps and reduce their size. The most often used pooling method for reducing feature maps is max pooling. It provides an abstract of picture representation areas, preventing overfitting. Moreover, it reduces the quantity of parameters required, resulting in a reduction in computational expenses [35].

The batch normalization (BN) layer's function is to normalize the convolutional layer's output. By utilizing higher learning rates, this technique accelerates the training process. Furthermore, it stops the model's gradients from disappearing during backpropagation. Additionally, BN layers enhance the resilience of DL models against incorrect weight initialization [39].

A dropout layer is commonly employed to address the issue of overfitting in neural network models. This layer randomly removes neurons based on a predetermined strategy during the training process. The dropout rate parameter significantly regulates the quantity of dropped neurons, thus controlling the probability of neuron elimination [40].

The final layer of a CNN network, known as the fully connected (FC) layer, serves as a crucial classifier that connects the layers within the network and ultimately determines the classification outcome. Usually, the SoftMax (SM) function is used for normalization in this layer. To improve the accuracy of classification, Support Vector Machine (SVM) or Random Forest (RF) techniques may replace the FC layers [35].

Feature extraction is achieved using convolutional and pooling layers, while picture classification uses FC layers. CNN employs pooling to aggregate related local attributes into a single feature and local connections to find local features. The output for each input MRI image is computed using FC layers or other classifiers [35].

1-2-2. Evaluation parameters

Performance evaluation metrics include the F1 score, Area Under the Precision-Recall (PR) Curve or AUC, Accuracy (Acc), Precision (Pre), Recall (Rec), Support (Sup), Sensitivity (Sen), and Specificity (Spe). The accuracy is the percentage of accurate predictions among all assumptions, including true positive and true negative. Precision counts how many real positive outcomes are anticipated accurately. Recall is the percentage of data samples from a class of interest—the "positive class"— accurately identified by a machine learning model as belonging to the class overall. A commonly used metric in machine learning is the F1 score, which combines recall and accuracy. The AUC is a standard metric for expressing a model's overall performance. The term support refers to the number of times a particular class appears in a given dataset. Sensitivity is a metric that measures how well a model can accurately predict positive instances for all available categories. On the other hand, specificity is a metric that evaluates a model's ability to predict negative instances for all available categories [32].

## 2. Review of related studies

Magnetic resonance brain imaging is a valuable tool for assessment and training, and the Open Access Series of Imaging Studies (OASIS) is an excellent resource for these purposes. The most recent release of the OASIS, known as OASIS-3, grants researchers access to a comprehensive set of neuroimaging datasets. This dataset, consisting of 6400 images, provides long-term neuroimaging, medical, cognitive, and biological marker data for normal aging and AD. It is divided into four categories based on the severity of Alzheimer's: Non-Demented (ND), Very Mild Demented (VMD), Mild Demented (MD), and Moderated Demented (MOD) [21].

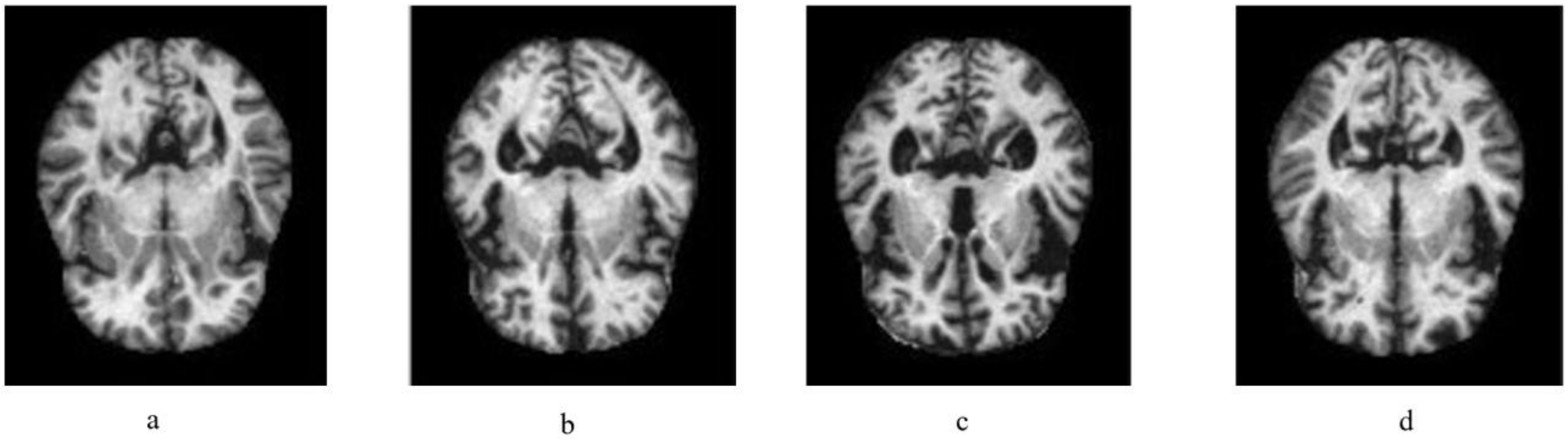

Figure 2. Instances of MR images for various stages of dement: (a) MD, (b) VMD, (c) MOD, and (d) ND [6].

### 2-1. Deep Triplet Networks

A deep triplet is a metric learning technique that employs distance measure to determine if the characteristics of images are similar and learn how to compare inputs. The loss term is vital in deep triplet networks (DTNs) as it establishes image differentiation. Research [22] shows that the triplet loss function is the most frequently used in DTNs [23]. Three samples are utilized to calculate a triplet loss: an anchor, a positive, and a negative data. The ultimate objective is to ensure that the distance between the anchor's features and the positive sample is less than between the anchor's features and the negative sample [22]. The model uses VGG16 as its basic network [24].

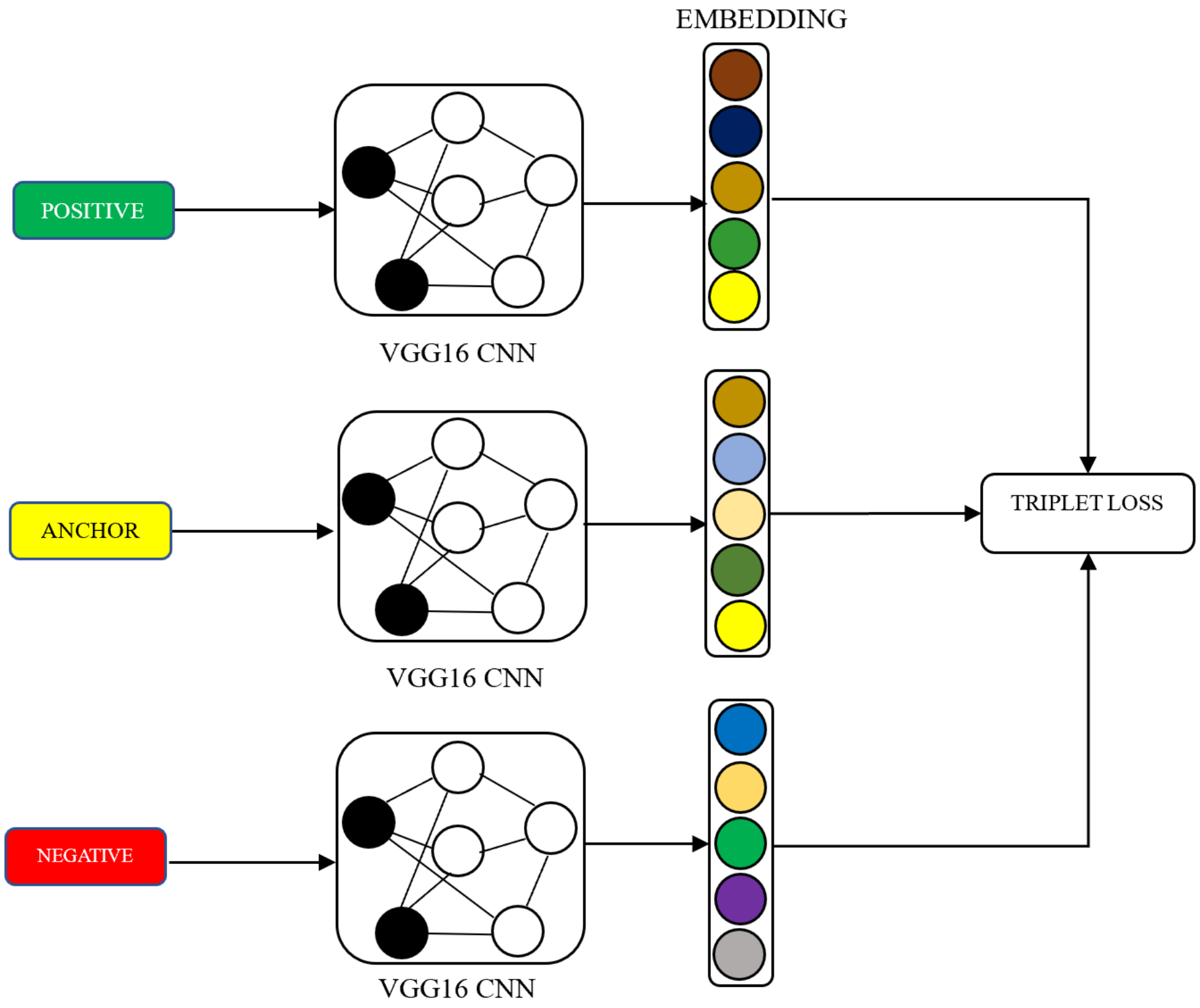

Figure. 3. Triplet Network [22]: Three networks share same structure and parameters.

Triplet loss raises the performance of the network [25]. The triplet loss receives three sample images as input - an anchor, a positive, and a negative. These are compared to the reference input (the anchor), a similar input (the positive), and a dissimilar input (the negative). While the negative class has a separate label, the positive samples and anchors are all the same class members. The network processes three samples, and the final layer determines the matching characteristics for each sample. Images belonging to the matching class should cluster together in the embedding space and be differentiated. The primary objective is to embed instances of the same label positioned closely in the embedding space while instances of separate labels are embedded far apart [22].

### 2-1-1. Brain MRI Image Recognition

The demographic of patients affected by AD ranges from 20 to 88 years of age. The suggested VGG-based model requires an image of a different size than the dataset images. The OASIS dataset is scaled using the picture dimension option [26].

As stated earlier, this network utilizes three input images to identify different types of AD. It generates corresponding feature embeddings for each input and calculates the distance between them. A "low" distance signifies that the images belong to a similar class, whereas a "high" distance indicates dissimilarity. A predetermined threshold must be established to aid the network in determining whether two inputs belong to the same class. If the distance between the features is lower than the threshold, the inputs are classified as belonging to the identical class [26].

### 2-1-2. Triplet Network Results

The OASIS open-access dataset was utilized for AD categorization, with ND, VMD, MD, and MOD being the four well-established classes. Table 1 presents the sample count for each class. To predict the multi-class classification results of diverse dementia phases, a conditional DTN and end-to-end learning model were engaged in this section. The suggested model achieved an impressive overall accuracy of 99.41% when applied to the selected MRI data from the OASIS dataset [26].

Table. 1. Summary of global clinical dementia rate [26].

| **Dementia Rate** | **No. of Samples** |
|---|---|
| ND | 167 |
| VMD | 87 |
| MD | 105 |
| MOD | 23 |

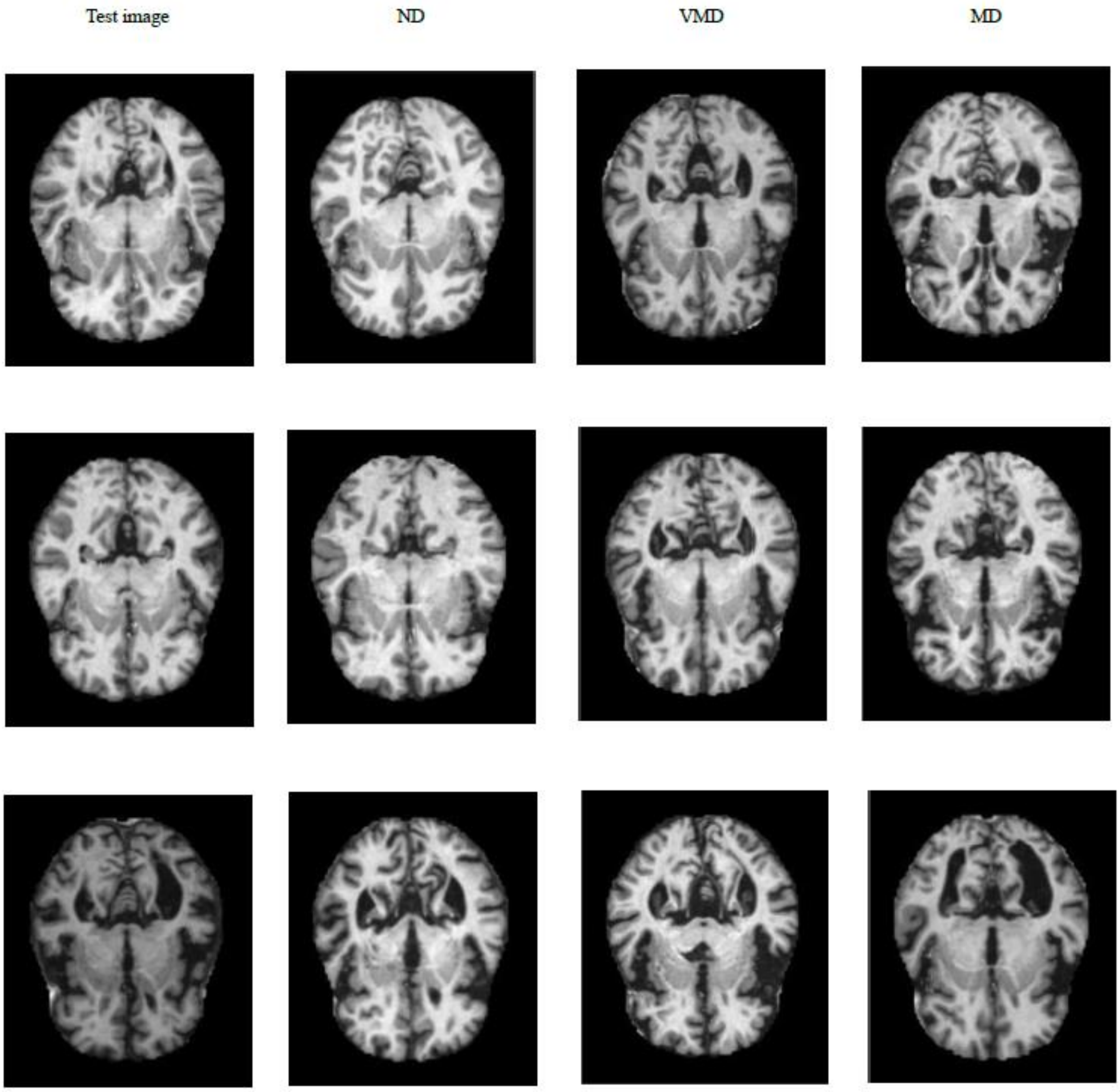

Figure. 4. ND, VMD, MD [26].

## 2-2. DEMNET

The CNN method is used for extracting the distinguishing characteristics, hence significantly increasing AD classification accuracy. Figure 5 represents the suggested DEMNET model process. The model comprises three basic steps: pre-processing, balancing the dataset with SMOTE, and classifying the data with DEMNET [27].

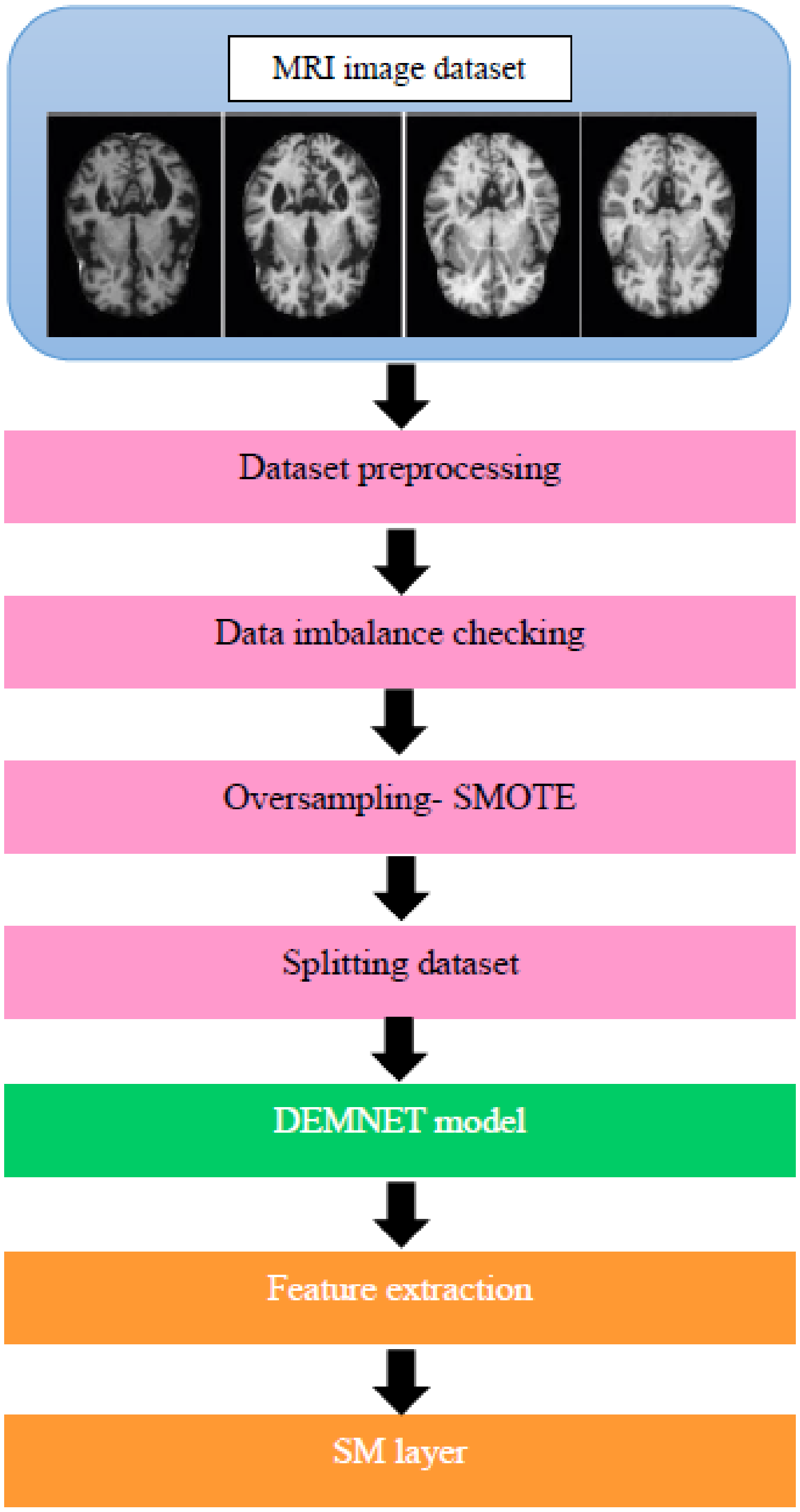

Figure. 5. DEMNET [27] model architecture for distinguishing phases of dementia.

The open-source site Kaggle provided the 6400 MR images for the AD dataset, separated into four classes: MD, MOD, ND, and VMD [27]. The four class example photos are displayed in Figure 6.

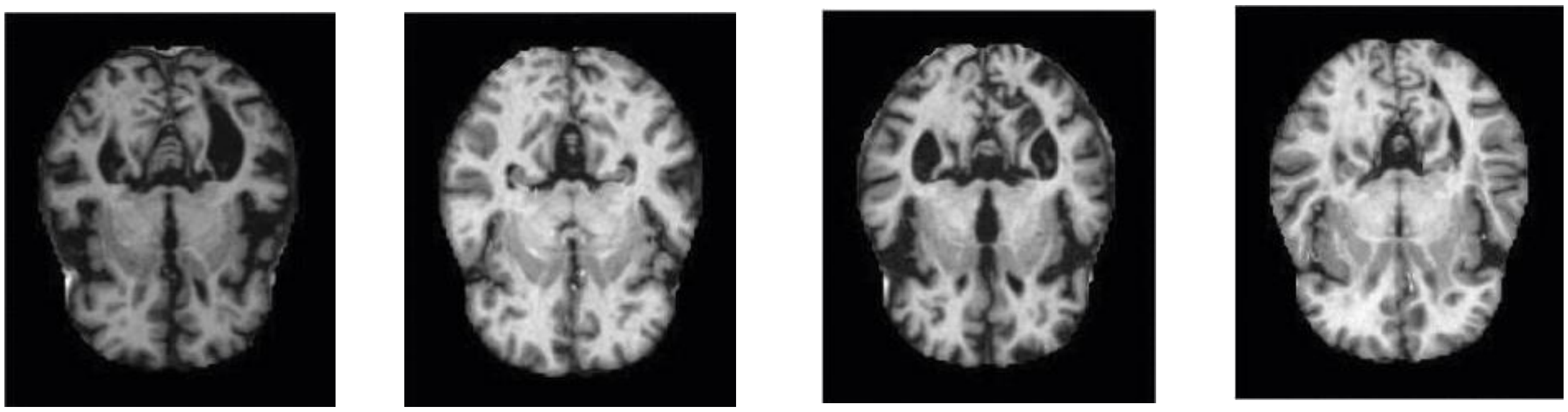

a b c d

Figure. 6. (a) MD, (b) VMD, (c) MOD, (d) ND [27].

Table 2 exhibits the dataset's distribution and the count of generated images, demonstrating the dataset's evident class imbalance. By randomly replicating minorities in the dataset to correspond to majorities, the SMOTE approach is employed to fix the class disparity issue in the dataset [28]. Utilizing SMOTE has benefits like limiting over-fitting and lessening the loss of information. The SMOTE technique increased the dataset distribution to 12800 photos, with 3200 images per class, as shown in Table 3 [27].

Table. 2. Dataset distribution [27].

| **Class** | **No. of Images** |
|---|---|
| ND | 2,240 |
| VMD | 64 |
| MD | 896 |
| MOD | 3,200 |

Table. 3. Dataset distribution after SMOTE [27].

| **Class** | **No. of Images** |
|---|---|
| ND | 3,200 |
| VMD | 3,200 |
| MD | 3,200 |
| MOD | 3,200 |

The dataset undergoes pre-processing and normalization before being input to a CNN, which effectively identifies the affected area of AD by extracting distinctive features. The CNN model is developed from scratch to categorize the dissimilar phases of dementia and identify AD accurately. The DEMENT architecture is composed of four DEMNET blocks, a Max pooling layer, two dropout layers, three dense layers, a SM activation layer, and two convolutional layers that utilize ReLU (Rectified Linear Unit) activation functions. ReLU is a widely utilized linear activation function for hidden layers [29].

DEMNET block comprises two convolutional layers with ReLU activation, a Max pooling layer, and a BN layer. This block in the suggested model uses a variety of filters to extract the particular features needed to categorize the stages of AD. Input data is passed through the ReLU activation function, producing zero output for negative values and the original value for positive values [27].

This architecture's design can extract as many discriminative elements as possible to draw attention to any stages of dementia present in an image. The output of the layers before the convolutional layer is normalized using BN. The DEMNET block uses BN, a regularization approach, to lessen overfitting in the suggested model [27].

After the convolutional layers, the flatten layer reduces the multi-dimensional input data into a single-column vector. The data outputted by the flatten layer is transferred as the input to the dense layer. The artificial neural networks (ANN) artificial dense layer carries out the same mathematical procedures. Three dense layers are employed, and each layer's neurons are linked to the layers' neurons in the subsequent levels. The SM function is utilized subsequent to the dense layer, wherein the count of neurons is equivalent to the count of classes [30]. The architecture of the suggested DEMNET for categorizing the phases of dementia is shown in Figure 7.

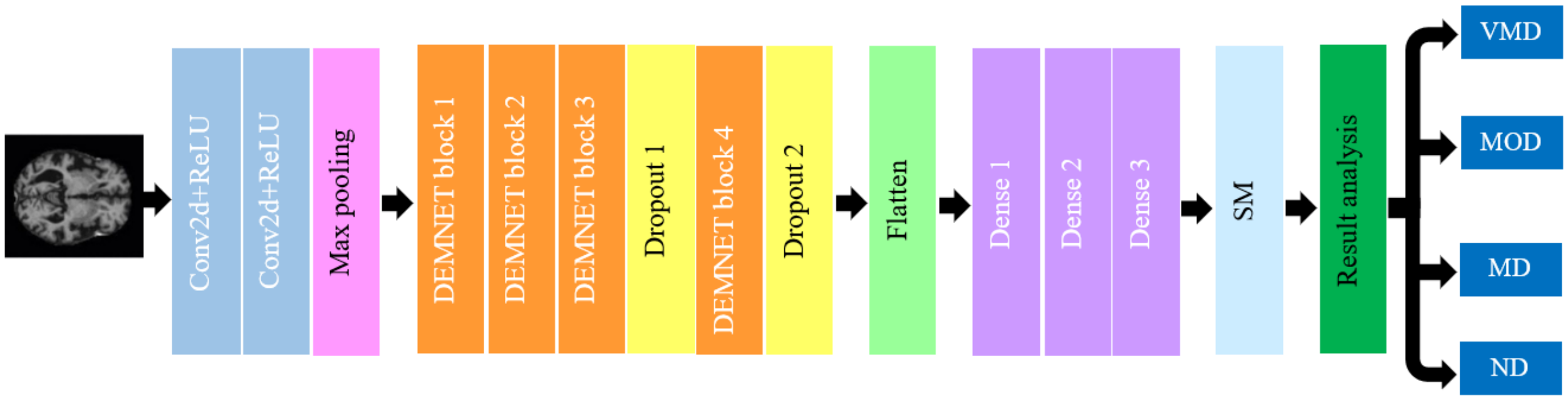

Figure. 7. Structure of the DEMNET [27] model for classifying stages of dementia.

2-2-1. DEMNET Results

A scenario without SMOTE and another with SMOTE were exerted to train the model. The AUC for each epoch is computed to evaluate the model's capability to differentiate between positive and negative classes. As a result of the imbalanced class size and overfitting issue, the model achieved a training accuracy of 96% but only 78% on validation without SMOTE. The SMOTE approach model has an overall 99% training accuracy and a 94% validation accuracy. The calculation resulted in 326 images from ND, 309 from VMD, 329 from MD, and 316 from MOD [27]. The ADNI dataset, which includes the five classes of AD (AD, MCI, EMCI, LMCI, and NC), was used for the experiment [31]. The ADNI dataset's 1296 images underwent resizing to suit the DEMNET model, resulting in impressive accuracy and performance on the Kaggle dataset. This was conducted to evaluate the DEMNET's resilience on diverse AD MRI datasets. The model garnered an accuracy rate of 84.83% and an AUC score of 95.62% [27].

Table. 4. DEMNET results [27].

| **Dataset** | **Class** | **PR** | **Recall** | **F1-Score** | **No. of Samples** |
|---|---|---|---|---|---|
| Kaggle | ND | 0.98 | 0.96 | 0.97 | 326 |
| | VMD | 0.99 | 1.0 | 1.0 | 309 |
| | MD | 0.88 | 0.98 | 0.93 | 329 |
| | MOD | 0.98 | 0.87 | 0.92 | 316 |

### 2-3. Combination of LeNet and AlexNet

The LeNet architecture contains several layers, including the convolution, pooling, FC, and output layers. Specifically, the convolution layer is vital in extracting essential features from the input data. Through the analysis of smaller subsets of input, the convolution process is able to retain the connections between pixels, which ultimately enhances the overall accuracy of the model [32].

The AlexNet architecture differs significantly from LeNet despite sharing many of the same basic principles. One major difference is that AlexNet is considerably larger than LeNet. It consists of eight layers, including an output layer after two dense layers and five convolutional layers, whereas LeNet is much smaller. Additionally, AlexNet uses ReLU activation rather than sigmoid activation. This model disproved the old-fashioned machine vision paradigm by showing that learning-based features may perform better than manually constructed ones [32].

It is common practice to improve classification performance outcomes using an ensemble of various DL models [33, 34]. The authors propose a unique model that parallelizes the layer-by-layer combining of both models, building on the original LeNet and AlexNet structures as a foundation. In addition, since the brain is a very limited region of interest, various sizes of the convolutional kernels aid in model learning [35]. A trio of brief parallel filters was submitted for the big convolutional filters included in the original designs [32].

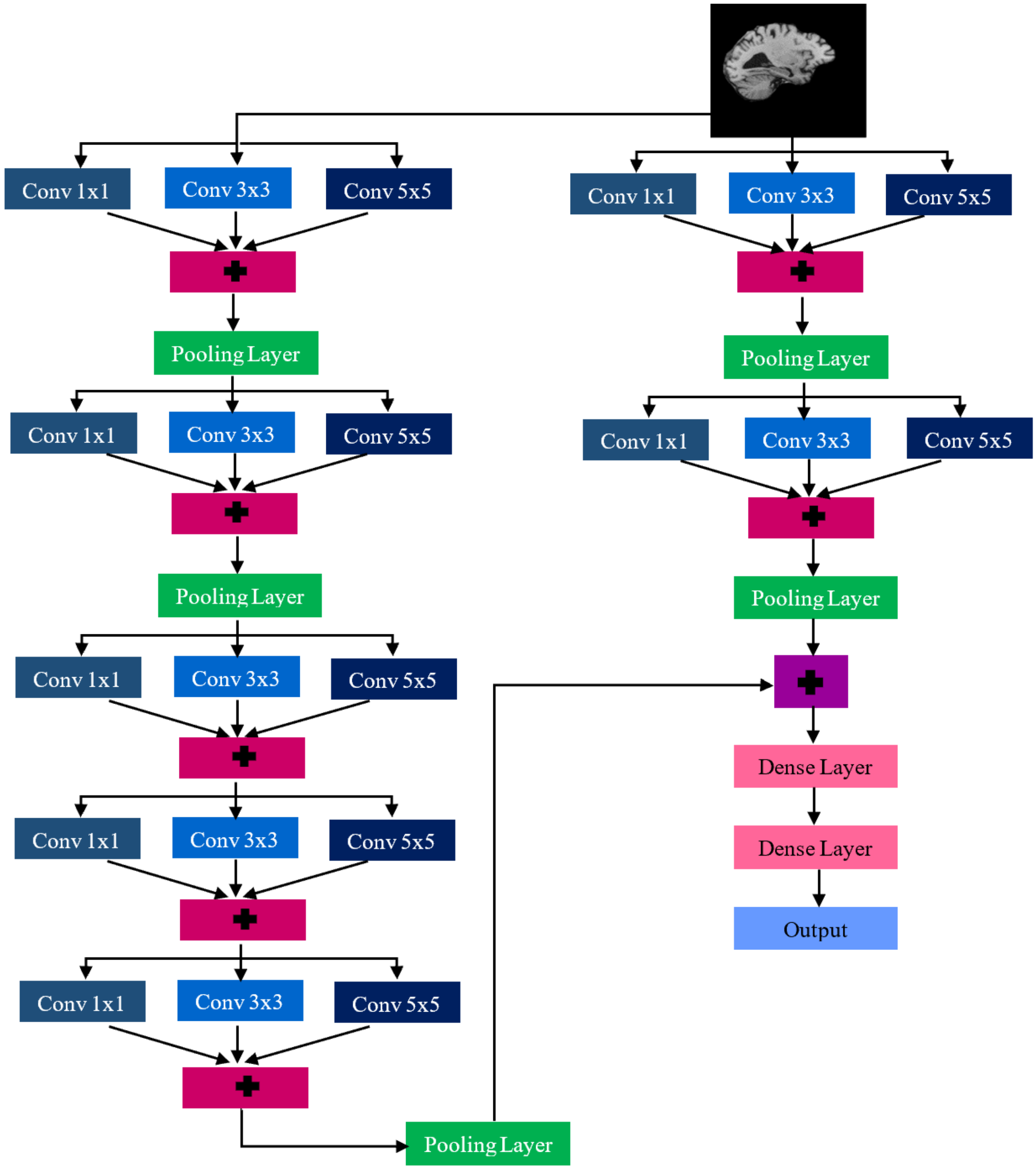

Figure. 8. Block diagram [32] of the model architecture.

Selecting fewer but more diverse parameters to accelerate the model is one of the main reasons for replacing regular convolution layers. It was divided into three different filter sizes rather than using the same enormous size over multiple kernels. This stage helped authors acquire many characteristics and reduced the number of variables, making the model operate more effectively [32]. The model's average performance is shown in Table 5.

Table. 5. Evaluating the model's efficiency for binary categorization [32].

| Dataset | Classes | Age | Pre | Rec | Acc | F1-Score | Average performance | Average time per epoch |
|---|---|---|---|---|---|---|---|---|
| ADNI | CN/MCI | 60–69 | 0.93 | 0.95 | 0.95 | 0.94 | 0.9358 | 72 s |
| | | 70–79 | 0.94 | 0.95 | 0.93 | 0.94 | | |
| | | 80+ | 0.94 | 0.92 | 0.95 | 0.94 | | |
| | MCI/AD | 60-69 | 0.93 | 0.92 | 0.93 | 0.96 | | |
| | | 70-79 | 0.93 | 0.93 | 0.96 | 0.96 | | |
| | | 80+ | 0.92 | 0.95 | 0.92 | 0.92 | | |
| | CN/AD | 60-69 | 0.96 | 0.94 | 0.96 | 0.93 | | |
| | | 70-79 | 0.93 | 0.92 | 0.92 | 0.93 | | |
| | | 80+ | 0.92 | 0.93 | 0.91 | 0.93 | | |

Table. 6. The proposed model's multi-class performance evaluation table [32].

| Dataset | Age | Pre | Rec | Acc | F1-Score |
|---|---|---|---|---|---|
| ADNI | 60–69 | 0.92 | 0.90 | 0.88 | 0.91 |
| | 70–79 | 0.88 | 0.89 | 0.83 | 0.89 |
| | 80+ | 0.85 | 0.84 | 0.83 | 0.85 |

Table 7. LeNet vs AlexNet vs proposed model [32].

| Model | Average performance | Average time per epoch |
|---|---|---|
| LeNet | 0.8025 | 68 s |
| AlexNet | 0.7150 | 79 s |
| LeNet + AlexNet | 0.9358 | 72 s |

### 2-3-1. Proposed model results

The intention of this research [32] was to efficiently and quickly classify AD using a hybrid approach that combines modified versions of LeNet and AlexNet. The model's binary class classification is evaluated by integrating modified versions of LeNet and AlexNet. The suggested approach generates noticeably fewer convolutional parameters than the existing one, which makes the model lighter and quicker. This work indicates that extracting features for DNN-based image classification does not always need huge convolutional filters. Combining multiple small kernels reduces parameters and processing time by extracting valuable attributes. It is crucial to highlight that the hybrid model surpasses the original AlexNet model in terms of speed, owing to its utilization of fewer convolutional parameters. Concerning execution efficiency and classification accuracy, the hybrid model stands out among all other models, as Table 7 exhibits. The hybrid approach thoroughly investigated multi-class classifications, emphasizing improved performance analysis through a 5-fold cross-validation strategy. Table 6 displays the multi-class classification performance and demonstrates that the suggested hybrid strategy may also be used successfully for multi-class categorization [32].

**2-4. ResNet-50**

The key objective of this investigation is to examine and optimize the classification efficacy of MRI in the prompt identification of AD through the utilization of DL and CNN methodologies. [35]. A CNN model for feature classification and extraction is designed and validated to accomplish this. The validated model is subjected to trials to evaluate its effectiveness in analyzing the characteristics of the FC layer of the CNN (ResNet). The outcomes are analyzed using three well-known traditional ML classifiers: SVM, RF, and SM. The proposed AD diagnostic approach involved several phases, starting with collecting MRI data. In the second phase, the MR images are pre-processed by resizing them to a size compatible with the CNN model. The pre-trained ResNet50 CNN is used for feature extraction to extract MRI image features, which are subsequently utilized for classification by three distinct classifiers: SM, SVM, and RF. The ResNet-50 architecture consists of an FC layer and five Conv block phases [36]. After data collection and image pre-processing, the dataset is split into three groups: a training set, a validation set, and a testing set. The CNN model is trained using a labeled dataset that enables the extraction of MRI feature vectors from the FC layer as part of the feature extraction process. The characteristic vectors are then subjected to each of the three separate classifiers. The model's fit to the unbiased training dataset is assessed using the validation set during model tuning [35]. To prevent the overfitting issue caused by the limited dataset, TensorFlow [37] and Keras [38] software are utilized to apply a pre-trained ResNet-50 CNN to the MRI scans as an alternative for creating a large dataset from scratch and training a CNN. ResNet-50 is considered the most innovative model in computer vision and DL, allowing for the training of subsequent layers with excellent performance.

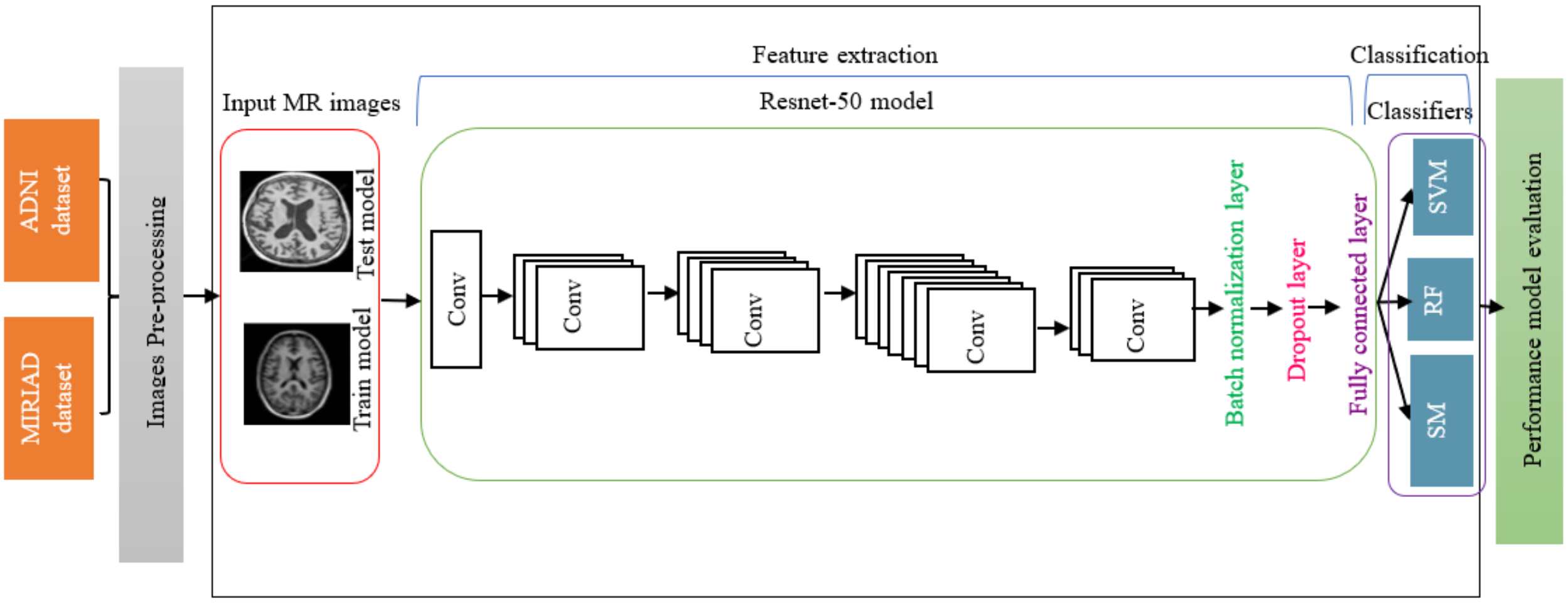

Figure. 9. The general structure of the model [35].

2-4-1. Results

The CNN model proposed in this study was developed based on the ResNet50 architecture, with specific modifications made to enhance its performance and mitigate overfitting. A BN layer was incorporated following each FC layer and the final convolution layer to regulate the output. Additionally, a dropout layer was inserted between the final FC layer and classifier to prevent the neural network from memorizing the training data too well, which can result in poor generalization performance on unobserved data. To evaluate the model's classification accuracy, SM, SVM, and RF classifiers were utilized on both the ADNI and MIRIAD datasets [35].

The primary aim of the experiment was to determine the optimal method for accurately identifying AD diagnostic pre-trained model ResNet50. SM is executed in the classifier layer for transfer learning on ResNet50. The model, which employs the ResNet50 architecture and SM, SVM, and RF algorithms, was tested on the ADNI and MIRIAD datasets. As per the results presented in Table 8, the SM-based model outperforms SVM and RF in all performance metrics for both datasets [35].

Table. 8. The three classifiers' performance in the suggested model [35].

| **Dataset** | **Classifier** | **Acc** | **F1-Score** |
|---|---|---|---|
| ADNI | SM | 99% | 98% |
| | SVM | 92% | 89% |
| | RF | 85.7% | 84% |
| MIRIAD | SM | 96% | 97% |
| | SVM | 90% | 87% |
| | RF | 84.8% | 79% |

The classifier's performance is assessed for each class [AD and NC].

Table. 9. The results from the Resnet50-SM experiment were conducted on the ADNI dataset [35].

| | **Pre** | **Rec** | **F1-score** | **Sup** |
|---|---|---|---|---|
| NC | 98% | 100% | 99% | 43 |
| AD | 100% | 97% | 98% | 32 |
| Acc | | | 99% | 75 |

Table. 10. Results of the Resnet50-SM test conducted on the MIRIAD [35].

| | **Pre** | **Rec** | **F1-score** | **Sup** |
|---|---|---|---|---|
| NC | 92% | 96% | 94% | 25 |
| AD | 98% | 96% | 97% | 48 |
| Acc | | | 96% | 73 |

The efficacy of the proposed diagnostic model for AD is evident from the high accuracy rates of AD classification, which were found to be 96.875% for the ADNI dataset and 95.83% for the MIRIAD dataset.

Furthermore, the results indicate that the three classifiers perform comparably. The SM classifiers have demonstrated the highest level of accuracy, providing strong evidence of their success. The RF classifier ranks third in accuracy, while the SVM classifier ranks second. The findings indicate that the dataset has no significant impact on the performance of the proposed model [35].

### 2-5. 3D DenseCNN using hippocampus MRI data

The measurement of hippocampal shrinkage through MRI scans is a valuable tool for determining the stage of AD [41, 42]. To examine hippocampal characteristics in relation to AD diagnosis, researchers have analyzed structural MRIs [42, 43] and utilized visual characteristics in SVMs [44]. While MRI scans can provide insight into changes in hippocampal volume [45], volumetric analysis alone can be limited in its ability to identify morphological changes that occur during AD progression. Researchers have begun using new methods for modeling and describing shapes to gain a more comprehensive understanding. Shape descriptors that rely on the Laplace-Beltrami (LB) spectrum are particularly useful as they are isometry-invariant and do not require complex pre-processing levels such as mapping, registration, and alignment [46, 28]. A new 3D model, DenseCNN, has been developed for AD classification using hippocampal segmentation [47]. Compared to other DL models, DenseCNN has fewer convolutional kernels and a simpler structure with fewer total parameters. However, recent research suggests deep convolutional neural networks (DCNN) may overlook global object shape properties [48, 49]. While DCNNs can recognize local shape characteristics like edge segments and connections, they lack representation of overall global form factors [48]. To address this limitation, researchers have presented DenseCNN2 - a 3D lightweight DCNN that integrates global shape and visual hippocampal segmentation information to enhance AD categorization. DenseCNN2 employs hippocampal segmentations and their corresponding global shape representations, in contrast to DenseCNN [50].

The brain has left and right hippocampi. Hippmapp3r, a segmentation method that utilizes 3D CNNs, separated the left and right hippocampi. It is known to be effective for MRI scans that involve brain atrophy and neurodegeneration. When compared to other segmentation algorithms, Hippmapp3r produces accurate and speedy hippocampus segmentation [51]. Figure 10 [50] displays the hippocampus segmentation findings from the AD and CN groups.

#### 2-5-1. DenseCNN architecture

Recent development has led to DenseCNN, a DCNN model that uses portions of the hippocampus to classify AD [47]. This model comprises three dense layers that include two convolutional layers mixed with BN and ReLU activation layers. The input data is reduced by a max pooling layer following the transition layers. For the left and right hippocampal areas, DenseCNN utilizes two streams commencing with a 3D convolutional layer, followed by BN and ReLU activation layers, to extract fundamental image data. Each stream incorporates two significant blocks and a transition layer featuring 8 and 16 filters, respectively. Each stream ends with a Global Average Pooling (GAP) layer, which converts multi-dimensional image attributes into 1D information. The final prediction is made using SM and FC layers after the two streams are combined and a dropout layer is applied. The model's architecture is depicted in Figure 11, with the CNN features being analyzed after the last GAP layer. This model collected Deep visual characteristics for the left and right hippocampus.

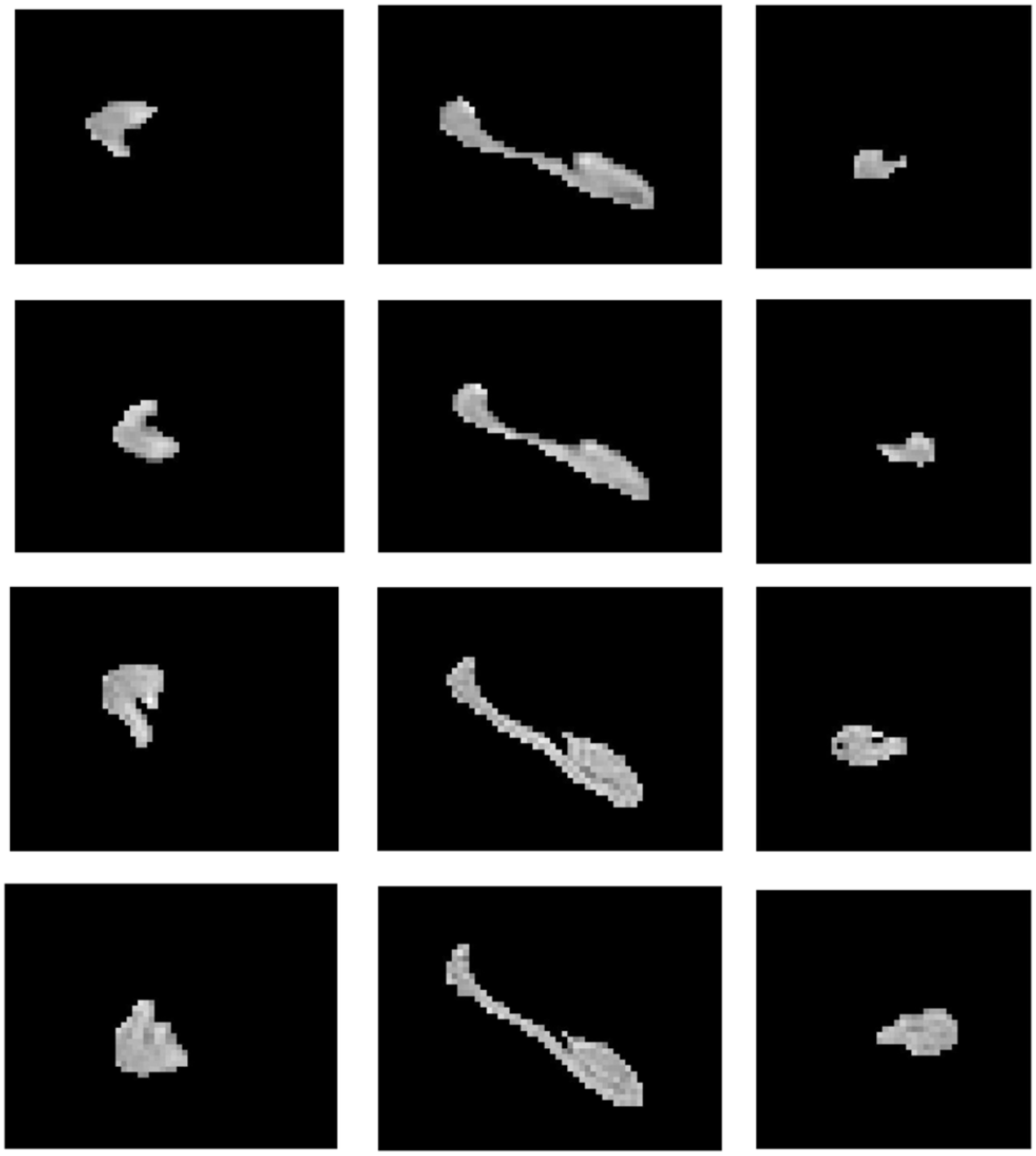

Figure. 10. Left and right hippocampus segmentation examples from AD and NC dataset of the ADNI [50].

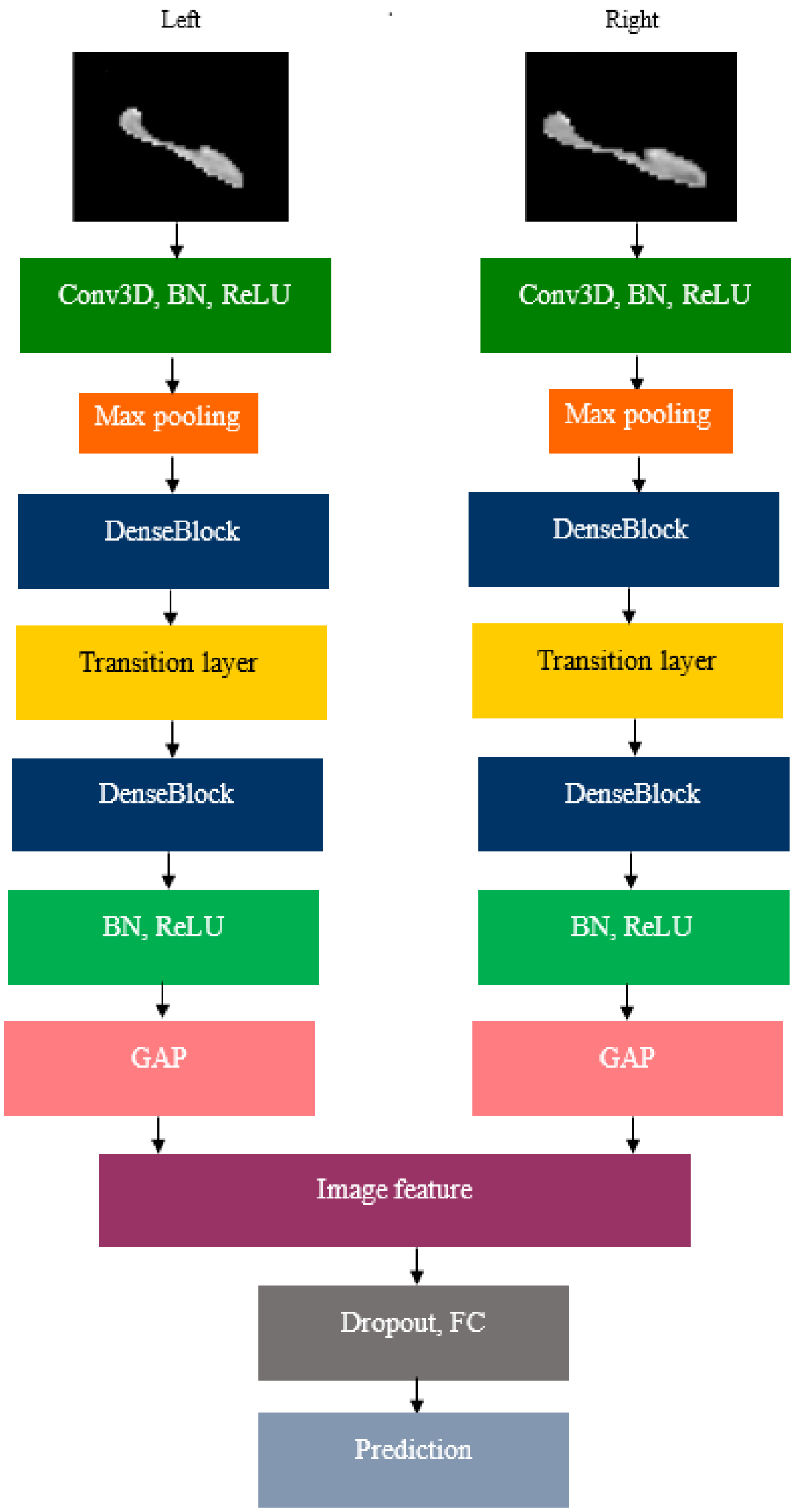

Figure. 11. The architecture of DenseCNN [50].

2-5-2. Using global shape features and DenseCNN features, simultaneously train DenseCNN2

The shape and DenseCNN characteristics were comprehensively developed and integrated into a neural network, utilizing a collaborative methodology. The LB spectrum provided a global description of the left and right hippocampi. A network structure comprising FC and SM layers was employed to classify AD to merge the shape features with the acquired attributes from DenseCNN [50].

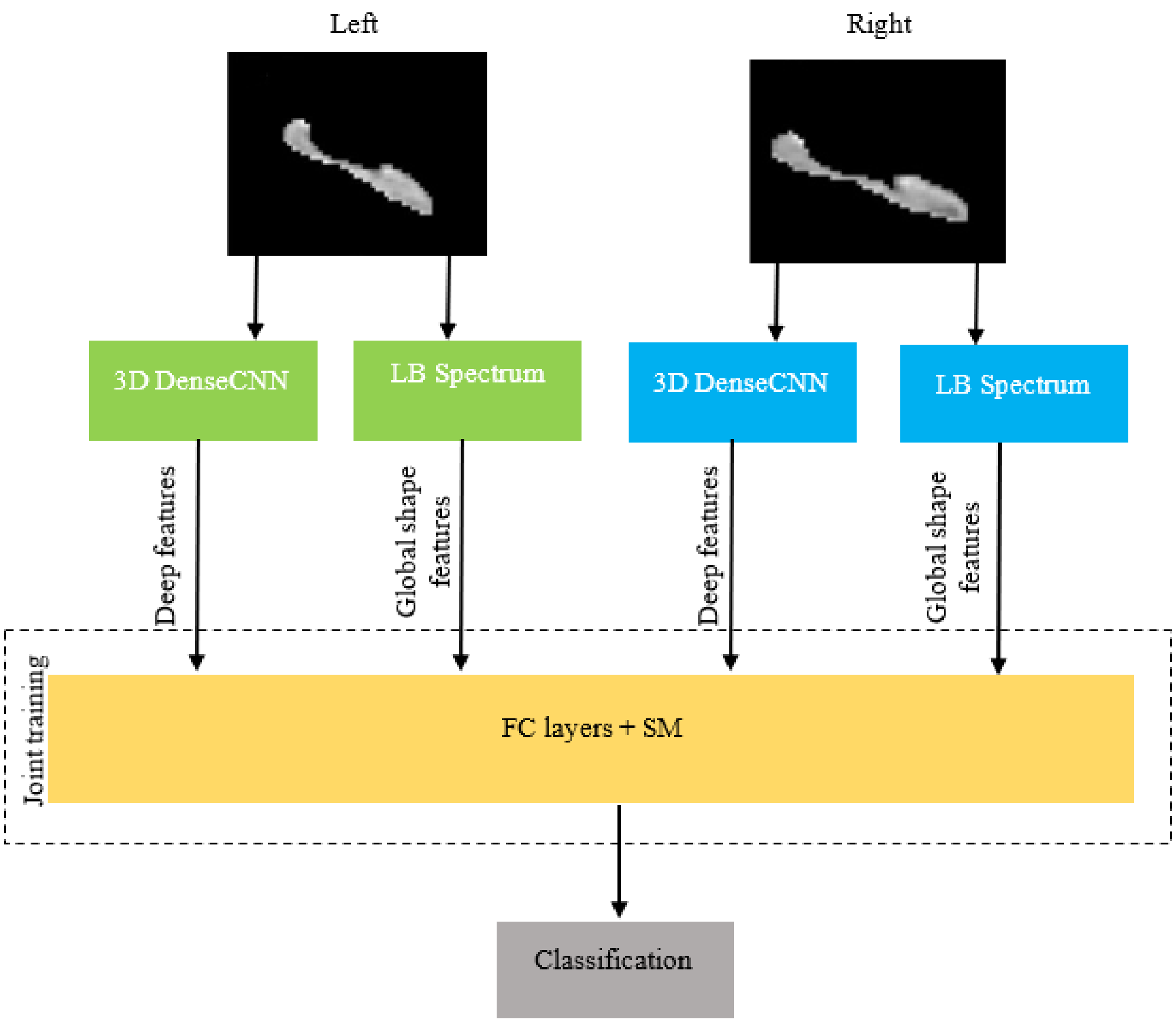

Figure. 12. The architecture of the joint model [50].

2-5-Results

2-5-3-1. Comparison of deep models utilizing either visual or shape features versus DenseCNN2 with combined visual and shape features

In Table 11, DenseCNN2 was compared to two other models: DL_shape, which uses only shape features, and DenseCNN with visual features. DL_shape was created with a network featuring all connected layers and SM layer classification. To train DenseCNN2, the network parameters were randomly initialized, and BN was used to learn deep features and speed up training while eliminating unimportant convolutional layer features. Dropout was also implemented to prevent overfitting. These findings suggest that combining CNN and form features in DenseCNN2 leads to better results for AD classification than using CNN features or shape features alone. The model has learned to use visual and overall form features of hippocampal segments to classify AD. Therefore, DenseCNN2 outperforms DenseCNN as it models visual and global shape features more accurately [50].

Table. 11. A comparative analysis of the efficacy of DL_shape, DenseCNN, and DenseCNN2 in the classification of AD vs. NC [50].

| **Method** | **Acc** | **AUC** |
|---|---|---|
| DL shape | 70.89 | 76.15 |
| DenseCNN | 89.91 | 96.42 |
| DenseCNN2 | 92.52 | 97.89 |

## 3. Discussion

This paper studied the correlation between AD and DL algorithms during the years 2021 to 2023. Our paper specifically and deeply focuses on five surveys that have demonstrated promising findings in AD prediction and diagnosis. By analyzing MRI scan datasets, these methodologies have the potential to provide valuable insights for healthcare professionals, thus aiding in informed medical decision-making and devising effective treatment plans for patients.

The predictive models presented in this paper have yielded impressive accuracies of 99.41%, 85%, 96%, 99% and 92% when utilized to diagnose and identify AD phases. These results highlight the considerable impact that DL methods can have on medical and early treatment.

Continued advancements in DL and artificial intelligence within the clinical domain necessitate further research and experimentation to attain superior and contemporary outcomes that positively influence human life.

Table. 12. Comparison of the mentioned studies with other state-of-the-art techniques.

| References | Dataset | | | Methodology | | Result |
|---|---|---|---|---|---|---|
| | Name | Modality | Size | Feature selection | Classifier | |
| Zhang et al. [52] | ADNI | MRI (PET) | 807 subjects<br>186 AD<br>226 CN<br>395 MCI | VGGNet-19 | SM | Overall Acc: 88.58%<br>AD vs. CN:<br>Acc: 95.12%<br>Spe: 93.97%<br>Sen: 95.94%<br>AUC: 0.9789%<br>AD vs MCI:<br>Acc: 82.4%%<br>Spe: 74.65%<br>Sen: 86.76%<br>AUC: 0.8720% |
| Liu et al. [53] | ADNI | MRI (PET) | 397 subjects<br>93 AD<br>76 pMCI<br>128 MCI<br>100 NC | 3DCNN | SM | AD vs. CN:<br>Acc: 84.97%<br>Spe: 87.37%<br>Sen: 82.65%<br>AUC: 90.63%<br>pMCI vs. NC:<br>Acc: 77.84%<br>Spe: 78.50%<br>Sen: 76.81%<br>AUC: 82.72%<br>sMCI vs. NC:<br>Acc: 60.09%<br>Spe: 54.21%<br>Sen: 65.29%<br>AUC: 62.38% |
| E. Jabason et al. [54] | OASIS | MRI | - | ResNet50 + DenseNet-201 | SM | AD vs. NC vs. MCI:<br>Acc: 95.23% |
| Farooq et al. [55] | ADNI | MRI | 149 subjects<br>33 AD<br>22 LMCI<br>49 MCI<br>45 NC | GoogleNet | SM | Overall Acc: 98.88<br>AD:<br>Spe: 99.2%<br>Sen: 97.9%<br>LMCI:<br>Spe: 99.9%<br>Sen: 99.9%<br>MCI:<br>Spe: 99.6%<br>Sen: 97.9%<br>NC:<br>Spe: 98.6%<br>Sen: 97.2% |
| Suriya et al. [27] | Kaggle | MRI | Dataset distribution:<br>3200 MD<br>3200 MOD<br>3200 ND<br>3200 VMD | DEMNET (with SMOTE) | SM | MD vs. MOD vs. ND vs. VMD:<br>Acc: 95.23% |
| Suriya et al. [27] | Kaggle | MRI | Dataset distribution:<br>896 MD<br>3200 MOD | DEMNET (without SMOTE) | SM | MD vs. MOD vs. ND vs. VMD:<br>Acc: 85% |
| | | | 2240 ND<br>64 VMD | | | |
| Orouskhani et al. [26] | OASIS | MRI | - | VGG-16 | Triplet loss | MOD vs VMD vs. MD vs. ND:<br>Acc: 99.41% |
| Hazarika et al. [32] | ADNI | MRI | - | LeNet + AlexNet | Dense layer | Average performance: 93%<br>CN/MCI:<br>Acc: 95%<br>MCI vs AD:<br>Acc: 96%<br>NC vs. AD:<br>Acc: 96% |
| D. AlSaeed et al. [35] | ADNI | MRI | 741 subjects<br>427 AD<br>314 NC | ResNet-50 | SM<br>RF<br>SVM | SM<br>Acc: 99%<br>Spe: 98%<br>Sen:99%<br>RF<br>Acc: 85.7%<br>Spe: 88%<br>Sen: 79%<br>SVM<br>Acc: 92%<br>Spe: 91%<br>Sen: 87% |
| D. AlSaeed et al. [35] | MIRIAD | MRI | 708 subjects<br>466 AD<br>243 NC | ResNet-50 | SM<br>RF<br>SVM | SM<br>Acc: 96%<br>Spe: 95%<br>Sen: 96%<br>RF<br>Acc: 84.8%<br>Spe: 84%<br>Sen: 73%<br>SVM<br>Acc: 90%<br>Spey: 91%<br>Sen: 87% |
| Ji et al. [56] | ADNI | MRI | - | ResNet50 + NASNet + MobileNet | SM | AD vs NC:<br>Acc: 98.59%<br>Spe: 100%<br>Sen: 97.29%<br>AD vs. MCI:<br>Acc: 97.65%<br>Spe: 100%<br>Sen: 96%<br>MCI vs NC:<br>Acc: 88.37%<br>Spe: 94%<br>Sen: 80.56% |
| Katabathula et al. [50] | ADNI | MRI | 933 subjects<br>326 AD<br>607 NC | 3D DenseCNN | SM | AD vs NC:<br>Acc: 92.52 %<br>Spe: 94.95 %<br>Sen: 88.20% |

## Abbreviations

| | |
|---|---|
| AD | Alzheimer’s Disease |
| MCI | Mild Cognitive Impairment |
| CN | Cognitive Normal |
| CNN | Convolutional Neural Network |
| DL | Deep Learning |
| ML | Machine Learning |
| pMCI | Progressive MCI |
| MCI | Stable MCI |
| EMCI | Early MCI |
| LMCI | Late MCI |
| CAD | Computer-Aided Diagnostic |
| OASIS | Open Access Series of Imaging Studies |
| ND | Non-Demented |
| VMD | Very Mild Demented |
| MD | Mild Demented |
| MOD | Moderated Demented |
| ANN | Artificial Neural Network |
| ADNI | Alzheimer’s Disease Neuroimaging Initiative |
| FC | Fully Connected Layers |
| VGG | Visual Geometry Group |
| MRI | Magnetic Resonance Imaging |
| SVM | Support Vector Machine |
| RF | Random Forest |
| MIRIAD | Minimal Interval Resonance in Alzheimer’s Disease |
| DCNN | Deep Convolutional Neural Network |
| BN | Batch Normalization |
| GAP | Global average pooling |
| Acc | Accuracy |
| Pre | Precision |
| Rec | Recall |
| Sup | Support |
| Sen | Sensitivity |
| Spe | Specificity |
| AUC | Area under curve |
| DTN | Deep triplet network |
| SM | SoftMax |
| NC | Normal cognitive |
| ReLU | Rectified linear unit |

## I. Declarations

### Availability of Data and Materials

Available.

### Funding

Not Applicable.

### Acknowledgments

Not Applicable.

### Conflict of Interest

- All authors have participated in (a) conception and design, or analysis and interpretation of the data; (b) drafting the article or revising it critically for important intellectual content; and (c) approval of the final version.

- This manuscript has not been submitted to, nor is it under review at, another journal or other publishing venue.

- The authors have no affiliation with any organization with a direct or indirect financial interest in the subject matter discussed in the manuscript

- The following authors have affiliations with organizations with direct or indirect financial interest in the subject matter discussed in the manuscript:

## Authors Biography

**Sarasadat Foroughipoor**

Sarasadat Foroughipoor is pursuing her Bachelor's degree in Computer Engineering at the University of Science and Research in Tehran. She has always been interested in the applications of deep learning in medicine, particularly for treating and diagnosing chronic illnesses. She intends to conduct more study and research to improve people's lives.

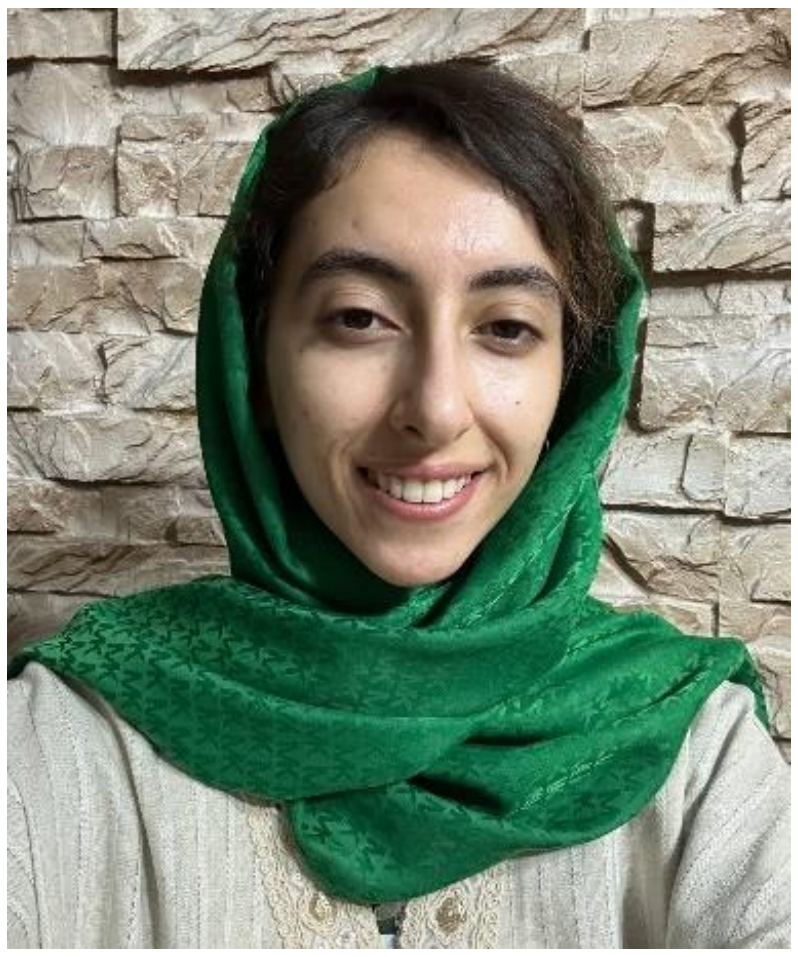

**Kimia Moradi**

Kimia Moradi is pursuing a Bachelor's degree in Computer Engineering at the Islamic Azad University Science and Research Branch. She has always been excited to work on projects that can help improve human lives. She wishes to be given more opportunities to contribute to the world.

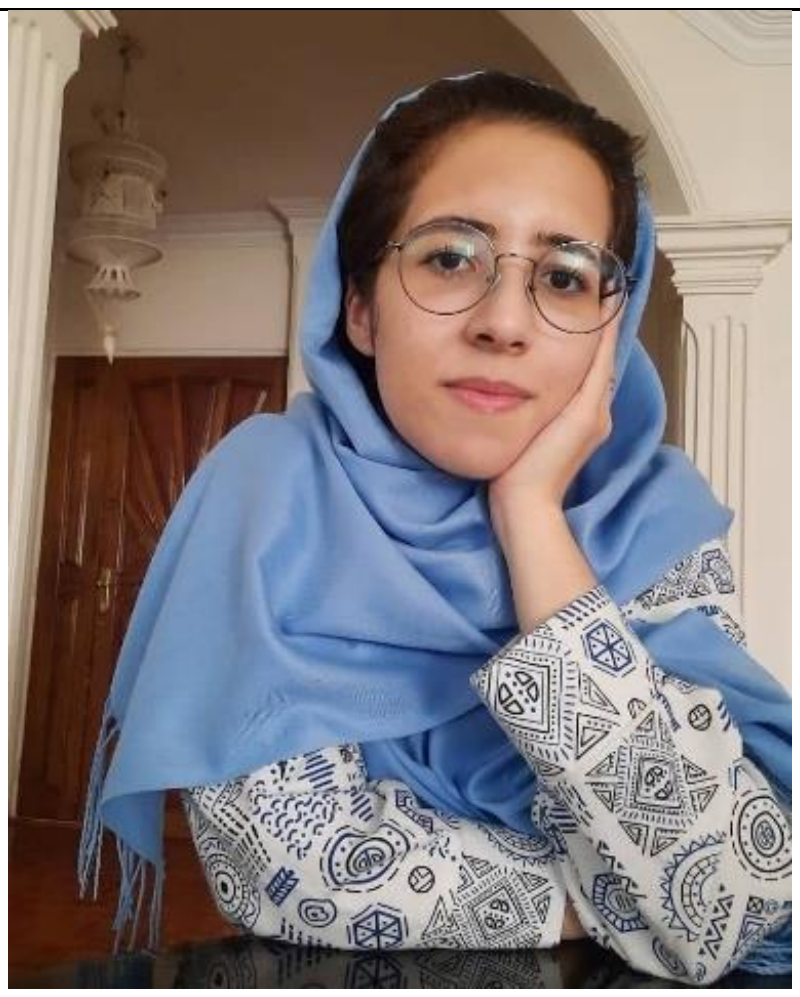

**Hamidreza Bolhasani, PhD**

AI/ML Researcher / Visiting Professor

Founder and Chief Data Scientist at DataBiox

Ph.D. Computer Engineering from Science and Research Branch, Islamic Azad University, Tehran, Iran. 2018-2023.

Fields of Interest: Machine Learning, Deep Learning, Neural Networks, Computer Architecture, Bioinformatics

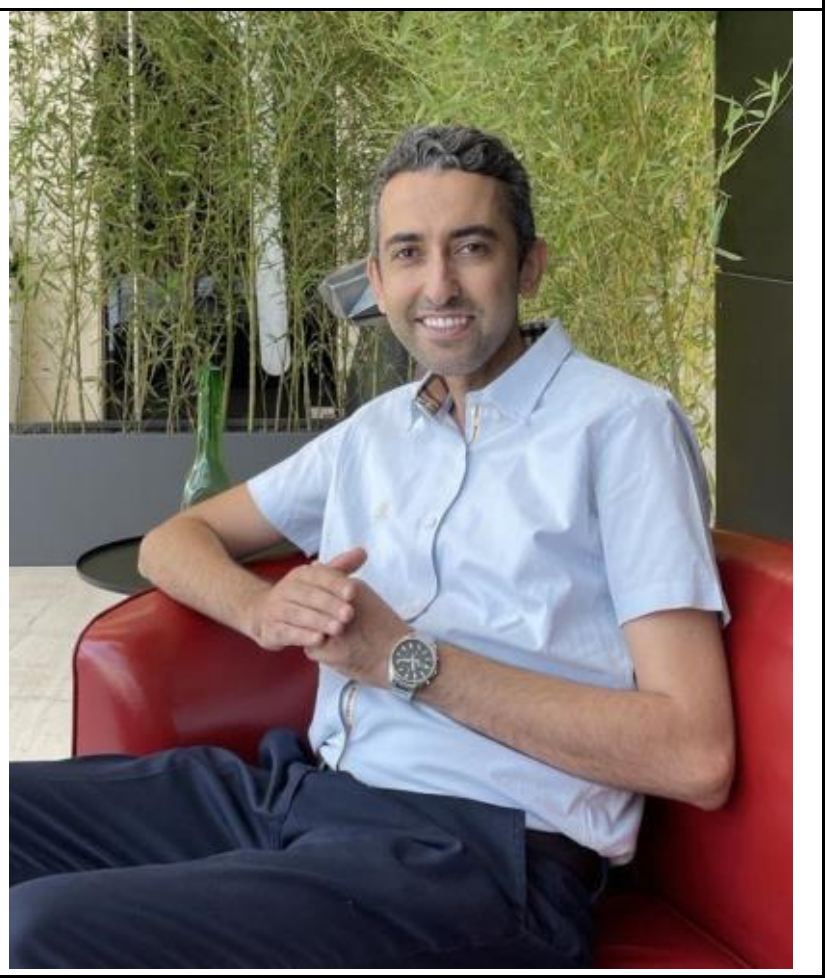